\documentclass[runningheads]{llncs}

\usepackage{graphicx}
\usepackage{longtable}
\usepackage{amsfonts}
\usepackage{amssymb}
\usepackage{bm}
\usepackage[utf8]{inputenc}
\usepackage[english]{babel}
\usepackage[fleqn]{amsmath} 
\usepackage{xstring}
\usepackage{setspace}
\usepackage{array}
\usepackage{subfig}
\usepackage{enumitem}
\usepackage{multirow}
\usepackage{algorithmic}
\usepackage{ctable} 

\newcommand{\FVar}[1] {\Hat{#1}} 
\newcommand{\ValRl}[1]{
    \IfEqCase{#1}{
        {x}{\alpha}
        {y}{\beta}
        {z}{\gamma}
        {w}{\delta}
        {v}{\zeta}
        {#1}{#1}
    }
}
\newcommand{\ValRlV}[1]{
    \IfEqCase{#1}{
        {x}{\bm{\alpha}}
        {y}{\bm{\beta}}
        {z}{\bm{\gamma}}
        {w}{\bm{\delta}}
        {v}{\bm{\zeta}}
        {#1}{\bm{#1}}
    }
}
\newcommand{\ValRp}[1]{
    \IfEqCase{#1}{
        {x}{\tilde{\alpha}}
        {y}{\tilde{\beta}}
        {z}{\tilde{\gamma}}
        {w}{\tilde{\delta}}
        {v}{\tilde{\zeta}}
        {#1}{\tilde{#1}}
    }
}
\newcommand{\ValRpV}[1]{
    \IfEqCase{#1}{
        {x}{\bm{\tilde{\alpha}}}
        {y}{\bm{\tilde{\beta}}}
        {z}{\bm{\tilde{\gamma}}}
        {w}{\bm{\tilde{\delta}}}
        {v}{\bm{\tilde{\zeta}}}
        {#1}{\bm{\tilde{#1}}}
    }
}
\newcommand{\FVal}[1]{
    \IfEqCase{#1}{
        {x}{\dot{\alpha}}
        {y}{\dot{\beta}}
        {z}{\dot{\gamma}}
        {w}{\dot{\delta}}
        {v}{\dot{\zeta}}
        {#1}{\dot{#1}}
    }
}
\newcommand{\FValV}[1]{
    \IfEqCase{#1}{
        {x}{\bm{\dot{\alpha}}}
        {y}{\bm{\dot{\beta}}}
        {z}{\bm{\dot{\gamma}}}
        {w}{\bm{\dot{\delta}}}
        {v}{\bm{\dot{\zeta}}}
        {#1}{\bm{\dot{#1}}}
    }
}

\newcommand{\Imply}{\Rightarrow}

\newcommand{\Sensor}[1]{
    \IfEqCase{#1}{
       {IW512}{L1}
       {IW576}{L2}
       {IW544}{L3}
       {IW554}{P1}
       {IW560}{P2}
        {#1}{#1}
    }
}

\newcommand{\TE}[1]{{\scriptsize\sffamily #1}} 
\newcolumntype{M}[1]{>{\centering\arraybackslash}m{#1}}
\newcolumntype{N}{@{}m{0pt}@{}}
\newcommand{\CompCode}[1]{{\small\sffamily #1}} 

\begin{document}
\raggedbottom
\title{Failure Mode Reasoning in Model Based Safety Analysis}
\author{H. Jahanian\inst{1} \and D. Parker\inst{2} \and M. Zeller \inst{3} \and A. McIver\inst{1} \and Y. Papadopoulos\inst{2}}
\institute{Macquarie University, Sydney, Australia\\ 
\email{hamid.jahanian@hdr.mq.edu.au; annabelle.mciver@mq.edu.au} \and
University of Hull, Hull, UK\\
\email{d.j.parker@hull.ac.uk; y.i.papadopoulos@hull.ac.uk} \and
Siemens AG, Munich, Germany\\
\email{marc.zeller@siemens.com}}
\maketitle

\begin{abstract}
Failure Mode Reasoning (FMR) is a novel approach for analyzing failure in a Safety Instrumented System (SIS). The method uses an automatic analysis of an SIS program to calculate potential failures in parts of the SIS. In this paper we use a case study from the power industry to demonstrate how FMR can be utilized in conjunction with other model-based safety analysis methods, such as HiP-HOPS and CFT, in order to achieve a comprehensive safety analysis of SIS. In this case study, FMR covers the analysis of SIS inputs while HiP-HOPS/CFT models the faults of logic solver and final elements. The SIS program is analyzed by FMR and the results are exported to HiP-HOPS/CFT via automated interfaces. The final outcome is the collective list of SIS failure modes along with their reliability measures. We present and review the results from both qualitative and quantitative perspectives. 

\keywords{FMR  \and HiP-HOPS \and CFT \and FTA.}
\end{abstract}

\section{Introduction}
In the process industry, Safety Instrumented Systems (SIS) are mechanisms that protect major hazard facilities against process-related accidents \cite{Ref_190}. Failure of SISs can result in catastrophic consequences such as loss of life and environmental damages. An SIS consists of hardware components and a software program. Failure Mode Reasoning (FMR) was introduced for calculating failure modes of SIS components based on an analysis of its program \cite{Ref_200}. Through a backward reasoning process on the SIS program, FMR calculates the SIS input failure modes that can result in a given undesired state at its output. Once the failure modes are identified, the probability of failure can be calculated too. 

Hierarchically Performed Hazard Origin \& Propagation Studies (HiP-HOPS) \cite{Ref_71} and Component Fault Trees (CFT) \cite{Ref_143} are two model-based dependability analysis techniques that can analyze failure modes of a system based on the failure behavior of its components. The failure models of components are combined to synthesize a system-level fault tree, which is then solved to generate qualitative and quantitative results. 

FMR was created to address a shortcoming in safety analyses in the process industry: the impact of SIS program. In this paper we demonstrate how other methods can achieve comprehensive failure analyses by employing FMR for an automatic analysis of the program. HiP-HOPS, for instance, offers automated synthesis and analysis of fault trees and FMEAs and state sensitive analysis of sequences, and it is also enriched with bio-inspired algorithms \cite{Ref_222}. However, the method still requires a first-pass manual annotation of failures, which is a challenging task when dealing with SIS programs. Likewise, CFT can benefit from an automated analysis of SIS programs conducted by FMR. In two independent experiments, we will integrate FMR with HiP-HOPS and CFT to analyze a case study from the power industry. Through qualitative and quantitative results we will show how such integrations can improve overall safety analysis. 

The rest of this paper is organized as follows: Section 2 provides an introduction to the underlying concepts of FMR and SIS failure analysis. Section 3 defines the case study and the method. Section 4 outlines the process of SIS input analysis in FMR. Sections 5 and 6 demonstrate the results of integrating FMR with HiP-HOPS and CFT. Section 7 discusses the challenges and achievements of the project, and section 8 wraps up the paper with a concluding note.

\section{SIS and FMR}\label{Sec_bkg}

A typical SIS consists of three main subsystems: sensors that measure the process conditions (e.g. pressure and temperature), logic solver (e.g. a CPU) that processes the program, and final elements (e.g. valves) that isolate the plant from a hazard when needed. The safety function achieved by a combination of sensors, logic solver and final elements to protect against a specific hazard is referred to as Safety Instrumented Function (SIF) \cite{Ref_190}.

As a layer of protection, the reliability of a SIF is commonly measured by its Probability of Failure on Demand (PFD): $PFD_{SIF} = PFD_{s}+PFD_{ls}+PFD_{fe}$; with $PFD_{s}$, $PFD_{ls}$ and $PFD_{fe}$ being the PFD of sensors, logic solver and final elements respectively, and $PFD_{SIF}$ the aggregated PFD of SIF \cite{Ref_190}. The PFD is calculated by using the failure rates of SIS components. A SIS component may fail in one of the following forms: Dangerous Detected (DD), Dangerous Undetected (DU), Safe Detected (SD) and Safe Undetected (SU) \cite{Ref_190}. A dangerous failure is a failure that prevents SIF from responding to a demand when a real hazard exists, and safe failure is the one that may result in a safety action being initiated by the SIF when there is no real hazard (i.e. Spurious Trip). The DU, DD, SU and SD elements are measured by failure rates $\lambda_{DU}$, $\lambda_{DD}$, $\lambda_{SU}$ and $\lambda_{SD}$. For a single component, the relationship between $\lambda_{DU}$ and the average PFD is expressed by $PFD_{avg}=\lambda_{DU}\tau/2$, in which $\tau$ is the Mission Time over which the average PFD is calculated. Other formulas are given by various sources to relate failure rates to the PFD and Spurious Trip Rate (STR) for general K-out-of-N (KooN) combinations \cite{Ref_188,Ref_196,Ref_197,Ref_194}.

Well established methods, such as Fault Tree Analysis (FTA) \cite{Ref_176,Ref_182} already exist in the industry for analyzing failure. FTA is a deductive method for failure analysis whereby a failure model, the fault tree, is analyzed to find the causes of a given undesired event. A fault tree is a graphical representation of failure, and it consists of events and logical gates that interconnect those events. The main outcome of an FTA is a set of minimal cut sets (MCS). An MCS is the smallest conjunction of a set of basic events that, together, can lead to the occurrence of the top event. Logically, MCSs represent AND combinations of basic events, and top event the OR combination of MCSs. Having the failure models and rates of occurrence for basic events one can calculate MCSs and the top event \cite{Ref_207,Ref_194,Ref_176}. 

With the growing complexity of industrial systems and availability of technology, FTA research has shifted towards modularization of models and automation of methods. HiP-HOPS and CFT are two examples of modular analysis of generic systems \cite{Ref_209,Ref_205}, as opposed to FMR which specializes in SIS programs. 

SIS programs are typically developed in graphical editors and in the form of Function Block Diagrams (FBD) \cite{Ref_203}. An FBD consists of standard Function Blocks (FBs) and their interconnections -- variables. Figure \ref{Fig_SIF_Config} includes a simplified picture of some FBs and their interconnections. As a more specific example, $y=(x_1+x_2)/2$ is an average value FB with output variable $y$ and input variables $x_1$ and $x_2$, which can connect this FB to other FBs in the program. Each FB, by itself, is fixed and known, but the function of the overall program depends on the selection of its constituting FBs and the way these FBs interact. Subsequently, the failure behavior of FBs can be defined independently, whereas the failure behavior of the program is identified based on its application-specific configuration. This is the underlying idea of FMR. In an automated process, the SIS program is scanned from its output towards its inputs as local failure behaviors are analyzed around each FB. The results of local analyses are then combined and simplified into a ``failure modes short list," which is also used for calculating SIS reliability measures \cite{Ref_200}.

FMR is based on a failure mode calculus. A failure mode is a manner in which the reported value of a variable in an SIS program deviates from its intended state; with the intended state being what the variable would read if SIS inputs were not affected by faults. Assuming that the SIS program is systematically correct, an undesired state at SIS output can only be caused by the propagation of input deviations through the program. FMR calculates the failure modes corresponding to such deviations by backward analysis of the program. The basic failure modes in FMR are expressed by $\FVal{h}$ and $\FVal{l}$ for real-valued variables, and $\FVal{t}$ and $\FVal{f}$ for Boolean variables. Here, $\FVal{h}$ is for \emph{false high}, $\FVal{l}$ for \emph{false low}, $\FVal{t}$ for \emph{false True} and $\FVal{f}$ for \emph{false False}. As an example, for the average value FB, $(\FVar{y}=\FVal{h})\Imply(\FVar{x_1}=\FVal{h}\vee\FVar{x_2}=\FVal{h})$ means if output $y$ is reading too high either input $x_1$ or $x_2$ must be too high. FMR combines such local reasoning statements, eliminates the intermediate variables, and produces a final, minimal statement comprising only SIS inputs and outputs.

FMR completes the SIS safety analysis by incorporating the functionally most important part of the system -- the program, and it does this by analyzing the actual program rather than a synthesized model. The process is automated and thus it saves time and effort, and offers accuracy and certainty.   

\section{Definition of the case}\label{Sec_Case}

Consider a gas-fired boiler with a high pressure drum for generating super-heat steam. The level and pressure in the drum are measured by three level transmitters and two pressure transmitters. Pressure measurement is used to modify the level readings: drum pressure can vary between 1 and 100 bars, causing wide-range changes to water density and thus to the level measurement. An SIS program uses thermodynamic calculations to correct the level readings based on pressure. Corrected level signals are compared to a preset threshold value, and if 2 out of 3 channels read extreme low, a trip is initiated at the outputs of the SIS logic solver to close the gas valves. Failing to shut the gas valves can result in excessive drum pressure, boiler tube rupture and eventually boiler explosion.

\begin{figure}[!ht]
\includegraphics[scale=0.52]{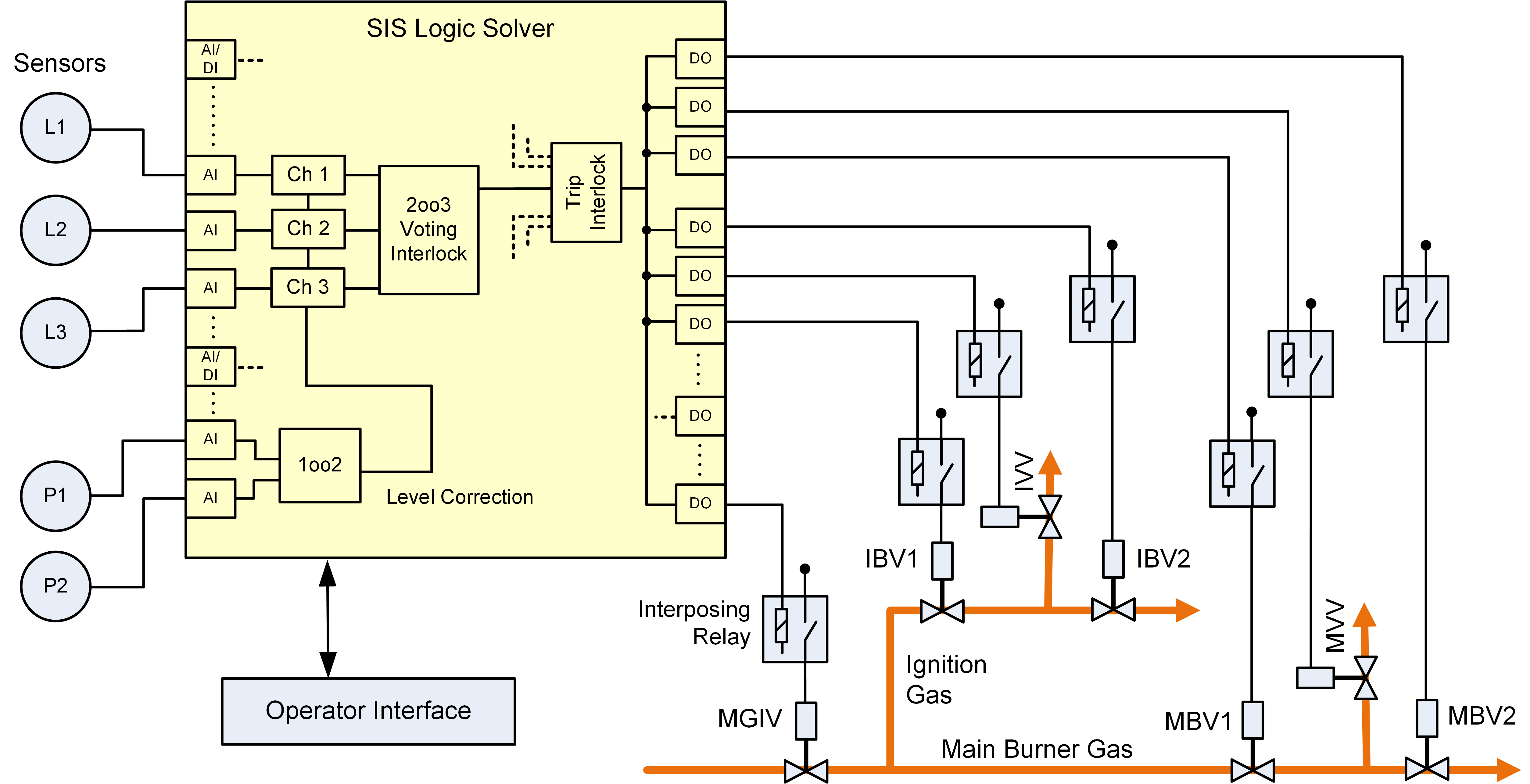}
\centering
\caption{SIS configuration}
\label{Fig_SIF_Config}
\end{figure}

As shown in Figure \ref{Fig_SIF_Config}, level transmitters L1-L3 and pressure transmitters P1-P2 are read in through analog input (AIs). The output of SIS program is connected to gas valves via output modules (DOs) and interposing relays. The gas skid consists of a main isolation valve (MGIV) and two sets of double-block-and-vent valves for the main burner (MBV1, MBV2, MVV) and ignition burner (IBV1, IBV2, IVV). During normal plant operation, MGIV and the block valves are open and the vent valves are closed. If a hazardous situation is detected, block valves should close and vent valves should open. MGIV is not considered a safety actuator and only closes during scheduled plant outages. 

The boiler is in its safe state (off) if both the main burner and ignition burners are shut. The following key failure states are defined:

\begin{itemize}
    \item The SIS is in a DU failure state if the level measurement fails to detect low drum level, or if the logic solver is not capable of responding to a detected low level, or if either the main burner valves (MBV1, MBV2) or ignition burner valves (IBV1, IBV2) are incapable of blocking the supply gas.
    \item The SIS will spuriously trip the boiler if MGIV, MBV1, MBV2, IBV1 or IBV2 closes when no real hazard exists. This may be due to a random failure of one of these valves or the failure of one of their upstream interposing relays, DO modules, CPU, AI modules or sensors.
\end{itemize}

To avoid the risk of extreme low drum level, the SIS program is designed to initiate a trip if any 2 out of 3 combination between low level and/or sensor fault is detected. Deviation between level signals alerts the operator but does not initiate a trip. Furthermore, pressure sensor P1 has priority over P2: if P1 is not detected faulty, the output of the 1oo2 block equals P1. 

Our objective in this case study is to analyze the SIF both qualitatively and quantitatively. We would like to determine the minimum combinations of component failures that can lead to SIS DU or ST failure. We would also like to calculate the likelihood of individual combinations and the aggregated PFD and STR. In the next three sections we will explain how we molded the SIS in FMR, HiP-HOPS and CFT. Independent from our case study models, we also created a reference fault tree in Isograph's FaultTree+ tool (\url{www.isograph.com}), of which no picture is shown here. The model was created to help us compare and evaluate the results of our analysis against one same, independent reference.

\section{Modeling SIS inputs in FMR}\label{Sec_model1}

This case study is based on a medium-scale power plant project where an SIS program performed 34 SIFs and included almost 100 hardwired inputs, 25 hardwired outputs and 250 software signals exchanged with an operator interface. The program comprised over 2170 function blocks with thousands of interconnections. A by-hand analysis of such programs would certainly be a challenge. Yet, in a typical large-scale power generation unit these figures may be five times greater, making a manual analysis almost impossible.

The input to the FMR tool is an offline copy of the entire SIS program. The analyst does not even need to know what the SIS program consists of. They only need to nominate a single variable in the program and the undesired state of that variable. The tool starts at the nominated point, traces the program backwards, and calculates the corresponding SIS input failure modes.  

\subsection{Qualitative analysis}\label{Sec_FMR_Qual}

We are interested in both DU and ST failure modes at the SIF output; i.e. at the final output of the Trip Interlock block in Figure \ref{Fig_SIF_Config}. The SIS is configured in a de-energize to trip setup. That is, a $False$ signal at the SIS output triggers a safety action and trips the plant. Thus, DU failure occurs when a real hazard exists but the SIS output is left $True$. Assuming that the SIS program is correct, a DU failure can only be due to the failure of SIS inputs in detecting the hazard. ST failure, on the other hand, occurs when the SIS output is set to $False$ due to safe failure of SIS inputs. In the FMR terminology, we are interested in SIF output being $\FVal{t}$ (for DU) and $\FVal{f}$ (for ST). 

A copy of the SIS program was imported to FMR, and the tag number and failure states of the SIF output were nominated. The tool analyzed the program and generated FM (failure mode) short lists shown in Tables \ref{Table_FMR_DU} and \ref{Table_FMR_ST}. 

\begin{longtable}[!h]{|M{0.7cm}|M{2.5cm}|M{2.5cm}|M{2.5cm}|M{2.5cm}|N|}
\hline 
\TE{1} & \TE{\Sensor{IW512}: healthy \& higher} & \TE{\Sensor{IW544}: healthy \& higher} &  & \\[-0.2em] 
\TE{2} & \TE{\Sensor{IW512}: healthy \& higher} & \TE{\Sensor{IW576}: healthy \& higher} &  & \\[-0.2em] 
\TE{3} & \TE{\Sensor{IW576}: healthy \& higher} & \TE{\Sensor{IW544}: healthy \& higher} &  & \\[-0.2em] 
\TE{4} & \TE{\Sensor{IW512}: healthy} & \TE{\Sensor{IW576}: healthy} & \TE{\Sensor{IW554}: healthy \& higher} & \\[-0.2em] 
\TE{5} & \TE{\Sensor{IW512}: healthy} & \TE{\Sensor{IW544}: healthy} & \TE{\Sensor{IW554}: healthy \& higher} & \\[-0.2em] 
\TE{6} & \TE{\Sensor{IW576}: healthy} & \TE{\Sensor{IW544}: healthy} & \TE{\Sensor{IW554}: healthy \& higher} & \\[-0.2em] 
\TE{7} & \TE{\Sensor{IW512}: healthy} & \TE{\Sensor{IW576}: healthy} & \TE{\Sensor{IW560}: healthy \& higher} & \TE{\Sensor{IW554}: faulty}\\[-0.2em] 
\TE{8} & \TE{\Sensor{IW512}: healthy} & \TE{\Sensor{IW544}: healthy} & \TE{\Sensor{IW560}: healthy \& higher} & \TE{\Sensor{IW554}: faulty}\\[-0.2em] 
\TE{9} & \TE{\Sensor{IW576}: healthy} & \TE{\Sensor{IW544}: healthy} & \TE{\Sensor{IW560}: healthy \& higher} & \TE{\Sensor{IW554}: faulty}\\ 
\hline 
\caption{FM short list for SIF output DU failure}\label{Table_FMR_DU}
\end{longtable}

\begin{longtable}[!h]{|M{0.7cm}|M{2.5cm}|M{2.5cm}|M{2.5cm}|M{2.5cm}|N|}
\hline 
\TE{1} & \TE{\Sensor{IW512}: healthy \& lower} & \TE{\Sensor{IW576}: healthy \& lower} &  & \\[-0.2em] 
\TE{2} & \TE{\Sensor{IW576}: healthy \& lower} & \TE{\Sensor{IW544}: healthy \& lower} &  & \\[-0.2em] 
\TE{3} & \TE{\Sensor{IW512}: healthy \& lower} & \TE{\Sensor{IW544}: healthy \& lower} &  & \\[-0.2em] 
\TE{4} & \TE{\Sensor{IW512}: faulty} & \TE{\Sensor{IW576}: healthy \& lower} &  & \\[-0.2em] 
\TE{5} & \TE{\Sensor{IW512}: faulty} & \TE{\Sensor{IW544}: healthy \& lower} &  & \\[-0.2em] 
\TE{6} & \TE{\Sensor{IW512}: healthy \& lower} & \TE{\Sensor{IW576}: faulty} &  & \\[-0.2em] 
\TE{7} & \TE{\Sensor{IW576}: faulty} & \TE{\Sensor{IW544}: healthy \& lower} &  & \\[-0.2em] 
\TE{8} & \TE{\Sensor{IW512}: faulty} & \TE{\Sensor{IW576}: faulty} &  & \\[-0.2em] 
\TE{9} & \TE{\Sensor{IW576}: healthy \& lower} & \TE{\Sensor{IW544}: faulty} &  & \\[-0.2em] 
\TE{10} & \TE{\Sensor{IW512}: healthy \& lower} & \TE{\Sensor{IW544}: faulty} &  & \\[-0.2em] 
\TE{11} & \TE{\Sensor{IW512}: faulty} & \TE{\Sensor{IW544}: faulty} &  & \\[-0.2em] 
\TE{12} & \TE{\Sensor{IW576}: faulty} & \TE{\Sensor{IW544}: faulty} &  & \\[-0.2em] 
\TE{13} & \TE{\Sensor{IW512}: healthy} & \TE{\Sensor{IW576}: faulty} & \TE{\Sensor{IW554}: healthy \& lower} & \\[-0.2em] 
\TE{14} & \TE{\Sensor{IW512}: healthy} & \TE{\Sensor{IW544}: faulty} & \TE{\Sensor{IW554}: healthy \& lower} & \\[-0.2em] 
\TE{15} & \TE{\Sensor{IW512}: faulty} & \TE{\Sensor{IW576}: healthy} & \TE{\Sensor{IW554}: healthy \& lower} & \\[-0.2em] 
\TE{16} & \TE{\Sensor{IW576}: healthy} & \TE{\Sensor{IW544}: faulty} & \TE{\Sensor{IW554}: healthy \& lower} & \\[-0.2em] 
\TE{17} & \TE{\Sensor{IW512}: healthy} & \TE{\Sensor{IW576}: healthy} & \TE{\Sensor{IW554}: healthy \& lower} & \\[-0.2em] 
\TE{18} & \TE{\Sensor{IW512}: faulty} & \TE{\Sensor{IW544}: healthy} & \TE{\Sensor{IW554}: healthy \& lower} & \\[-0.2em] 
\TE{19} & \TE{\Sensor{IW576}: faulty} & \TE{\Sensor{IW544}: healthy} & \TE{\Sensor{IW554}: healthy \& lower} & \\[-0.2em] 
\TE{20} & \TE{\Sensor{IW512}: healthy} & \TE{\Sensor{IW544}: healthy} & \TE{\Sensor{IW554}: healthy \& lower} & \\[-0.2em] 
\TE{21} & \TE{\Sensor{IW576}: healthy} & \TE{\Sensor{IW544}: healthy} & \TE{\Sensor{IW554}: healthy \& lower} & \\[-0.2em] 
\TE{22} & \TE{\Sensor{IW512}: healthy} & \TE{\Sensor{IW576}: faulty} & \TE{\Sensor{IW560}: healthy \& lower} & \TE{\Sensor{IW554}: faulty}\\[-0.2em] 
\TE{23} & \TE{\Sensor{IW512}: healthy} & \TE{\Sensor{IW544}: faulty} & \TE{\Sensor{IW560}: healthy \& lower} & \TE{\Sensor{IW554}: faulty}\\[-0.2em] 
\TE{24} & \TE{\Sensor{IW512}: faulty} & \TE{\Sensor{IW576}: healthy} & \TE{\Sensor{IW560}: healthy \& lower} & \TE{\Sensor{IW554}: faulty}\\[-0.2em] 
\TE{25} & \TE{\Sensor{IW576}: healthy} & \TE{\Sensor{IW544}: faulty} & \TE{\Sensor{IW560}: healthy \& lower} & \TE{\Sensor{IW554}: faulty}\\[-0.2em] 
\TE{26} & \TE{\Sensor{IW512}: healthy} & \TE{\Sensor{IW576}: healthy} & \TE{\Sensor{IW560}: healthy \& lower} & \TE{\Sensor{IW554}: faulty}\\[-0.2em] 
\TE{27} & \TE{\Sensor{IW512}: faulty} & \TE{\Sensor{IW544}: healthy} & \TE{\Sensor{IW560}: healthy \& lower} & \TE{\Sensor{IW554}: faulty}\\[-0.2em] 
\TE{28} & \TE{\Sensor{IW576}: faulty} & \TE{\Sensor{IW544}: healthy} & \TE{\Sensor{IW560}: healthy \& lower} & \TE{\Sensor{IW554}: faulty}\\[-0.2em] 
\TE{29} & \TE{\Sensor{IW512}: healthy} & \TE{\Sensor{IW544}: healthy} & \TE{\Sensor{IW560}: healthy \& lower} & \TE{\Sensor{IW554}: faulty}\\[-0.2em] 
\TE{30} & \TE{\Sensor{IW576}: healthy} & \TE{\Sensor{IW544}: healthy} & \TE{\Sensor{IW560}: healthy \& lower} & \TE{\Sensor{IW554}: faulty}\\ 
\hline 
\caption{FM short list for SIF output ST}\label{Table_FMR_ST}
\end{longtable}

Each row in Tables \ref{Table_FMR_DU} and \ref{Table_FMR_ST} represents an AND combination of input FMs that can result in the given output FM. A quick comparison with the description of the SIS program we described in Section \ref{Sec_Case} shows that FMR has identified failure modes as expected. In analyses where unexpected FMs are detected, engineers can use the information to correct or modify the program.
 
\subsection{Quantitative analysis}\label{Sec_FMR_Quan}

In the second stage, FMR performs a quantitative analysis to determine the probability of occurrence of failure. The FMR tool uses its internal project database to store failure data. In this database, each FM is described by a failure type and a likelihood value. The failure type can be ``Fixed" probability, failure-repair ``Rate" or ``Dormant". The likelihood value indicates the probability of failure (i.e. unavailability) or the frequency of occurrence (in a time interval).

A Fixed probability model is used when the occurrence of a basic event is expressed independently from time and the repair process. The unavailability ($q$) of a component with fixed probability value of $p$ will be: $q=p$.

The Rate model is suitable for repairable elements. These are the components for which the occurrence of a fault is detected and for which repair and restoration procedures are in place. The only time that the component is unavailable will be the time that it is under repair. The time interval is known as MTTR (Mean Time To Restoration) and the unavailability of such components will be \cite{Ref_194}:
\begin{equation}\label{Eq_Rateq}
q(t)=\lambda(1-e^{-(\lambda+\mu)t})/(\lambda+\mu)
\end{equation}    
with $\lambda$ being the failure rate and $\mu=1/MTTR$ the repair rate. These rates are often expressed \textit{per hour}. For a steady-state estimation of Eq. \ref{Eq_Rateq}, $t$ is assigned the constant value of Risk Assessment Time, often equal to Mission Time.

A Dormant model is used when a basic event represents the undetected fault of a component that undergoes periodic proof testing. Here we use \cite{Ref_194}:
\begin{equation}\label{Eq_Dormantq}
q=1 - (1-e^{-\lambda\tau})/(\lambda\tau)
\end{equation}    

The failure rates and models used in this project are listed below:

\begin{itemize}
\item A sensor being healthy \& higher (or healthy \& lower): $\lambda_{DU}=\lambda_{SU}=50~FIT$\footnote{$1\; FIT= 1\;in\; 10^9\; hours$}, $\tau=2~years$, and the event is modeled as Dormant. Reading high (or low) values without having an indication of fault is an undetected fault. This is why the Dormant model is selected for this type of failure. Depending on the direction of fault, the failure mode can be considered dangerous or safe. In this case study higher readings lead to DU failures and lower readings lead to ST; due to the intended functionality of the SIF.
\item A sensor having a detected fault: $\lambda_{DD}=\lambda_{SD}=250~FIT$, $MTTR=8~hours$, and the event is modeled as failure-repair Rate. 
\item A sensor being healthy: $q=0.999$, modeled as a Fixed probability value. It is assumed  that a transmitter is healthy for 99.9\% of time.
\end{itemize}

Basic events with fixed probability values cannot be expressed in frequency form. For the Rate and Dormant models, the frequency of a basic event will be:
\begin{equation}\label{Eq_BEFreq}
w=\lambda(1-q)
\end{equation}

Collective calculation of probability in FMR is similar to quantitative analysis of MCSs and top events in FTA. An MCS consists of one or several basic events, similar to one row in Tables \ref{Table_FMR_DU} and \ref{Table_FMR_ST}. With $Q_{MCS}$ and $W_{MCS}$ being the unavailability and frequency of an MCS with $n$ basic events:
\begin{equation}\label{Eq_MCSQW}
Q_{MCS}=\prod_{i=1}^{n}q_i \qquad \text{and}\qquad W_{MCS}=\sum_{i=1}^{n}w_i\prod_{\substack{j=1\\  j\neq i}}^{n}q_j
\end{equation}

The top event of a fault tree is an OR combination of its MCSs. The unavailability and frequency of the top event are approximated by:\footnote{Eq. \ref{Eq_MCSQW} is commonly referred to as Esary-Proschan method and is used by FTA tools such as FaultTree+, Arbor and Item. See \cite{Ref_207} for derivation of underlying concepts.}
\begin{equation}\label{Eq_TEQW}
Q_{TE}=(\prod_{i=1}^{c}q_i) (1-\prod_{k=1}^{m}(1-Q_k))\quad \text{and}\quad W_{TE}=\sum_{i=1}^{m}W_i\prod_{\substack{j=1 \\  j\neq i}}^{m}(1-Q_j)
\end{equation}

Here, $q_i$ is the unavailability of a basic event that is common between all MCSs, $c$ the number of common basic events, $Q_k$ the unavailability of the $k$th MCS excluding the common basic events, $Q_j$ the unavailability of the $j$th MCS, $W_i$ the frequency of the $i$th MCS, and $m$ the number of constituting MCSs.

Using Eqs. \ref{Eq_MCSQW} and \ref{Eq_TEQW}, FMR generated the following results for our case study. The results were verified by replicating the models in FaultTree+, which showed no differences.

\begin{itemize}
\item Aggregated unavailability for DU mode: $Q_{DU}=1.31E-03$, consisting of:
	\begin{itemize}
	\item FMs in rows 1-3 of Table \ref{Table_FMR_DU}, each with $Q_{FM}=1.92E-07$. 
	\item FMs in rows 4-6 of Table \ref{Table_FMR_DU}, each with $Q_{FM}=4.37E-04$. 
	\item FMs in rows 7-9 of Table \ref{Table_FMR_DU}, each with $Q_{FM}=8.75E-10$. 
	\end{itemize}
\item Aggregated frequency for ST mode: $W_{ST}= 1.33E-03~p.h.$, with  $W_{FM}=50~FIT$ for rows 17, 20 and 21 of Table \ref{Table_FMR_ST}, and $W_{FM}=0.0$ for other rows. 
\end{itemize}

\section{Integration with HiP-HOPS}\label{Sec_model2}
There are three phases to the analysis process in HiP-HOPS: modeling, synthesis, and analysis \cite{Ref_209}. In the manual modeling phase, a topological model of the system is created that details the components of the system and indicates how the components are connected together to allow the flow of data. Components can be grouped together hierarchically in sub-systems to help manage the complexity and allowing for refinement of the model as the design progresses. The components of the model are then augmented with local failure behavior that defines how each component's output can deviate from its normal expected behavior. The failure logic further documents how these output deviations can be caused by the combination of internal failure modes of the component and/or the propagation of deviations of the inputs of the component.

The second HiP-HOPS phase is the automatic synthesis of an interconnected set of fault trees that are produced by traversing the model of the system from its outputs to its inputs. It is during this phase that the failure logic defined in the modeling phase is combined by following the connections between the ports of the components and matching previously unrealized input deviations with output deviations of the same class that trigger them. This results in a model of the propagation of failure throughout the system.

The final stage is the analysis of the interconnected fault trees generated during synthesis. This begins with a qualitative pass that contracts the fault trees and removes the redundant logic resulting in the MCSs. The MCSs are then used together with the failure models of the components to run the quantitative pass and produce system unavailability and failure frequency measures.

\begin{figure}[!ht]
\centering
\subfloat[system level\label{Fig_MATLAB}]{\includegraphics[width=0.7\columnwidth]{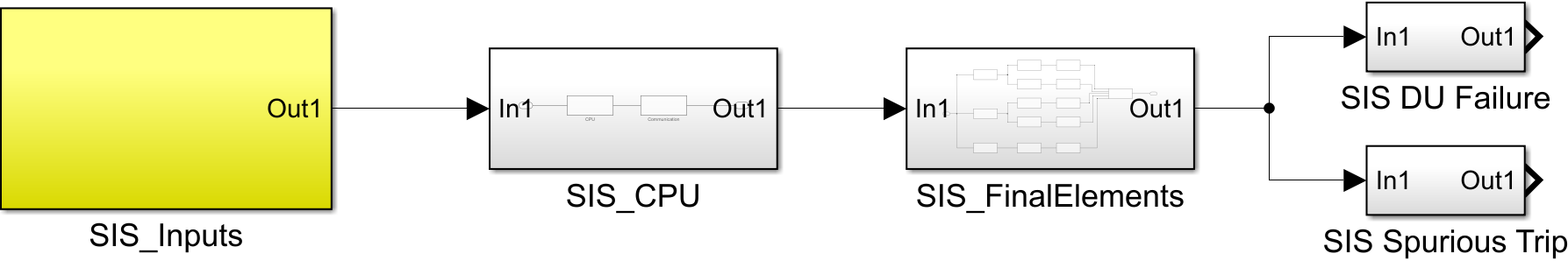}}

\subfloat[final elements\label{Fig_MATLAB1}]{\includegraphics[clip,width=0.7\columnwidth]{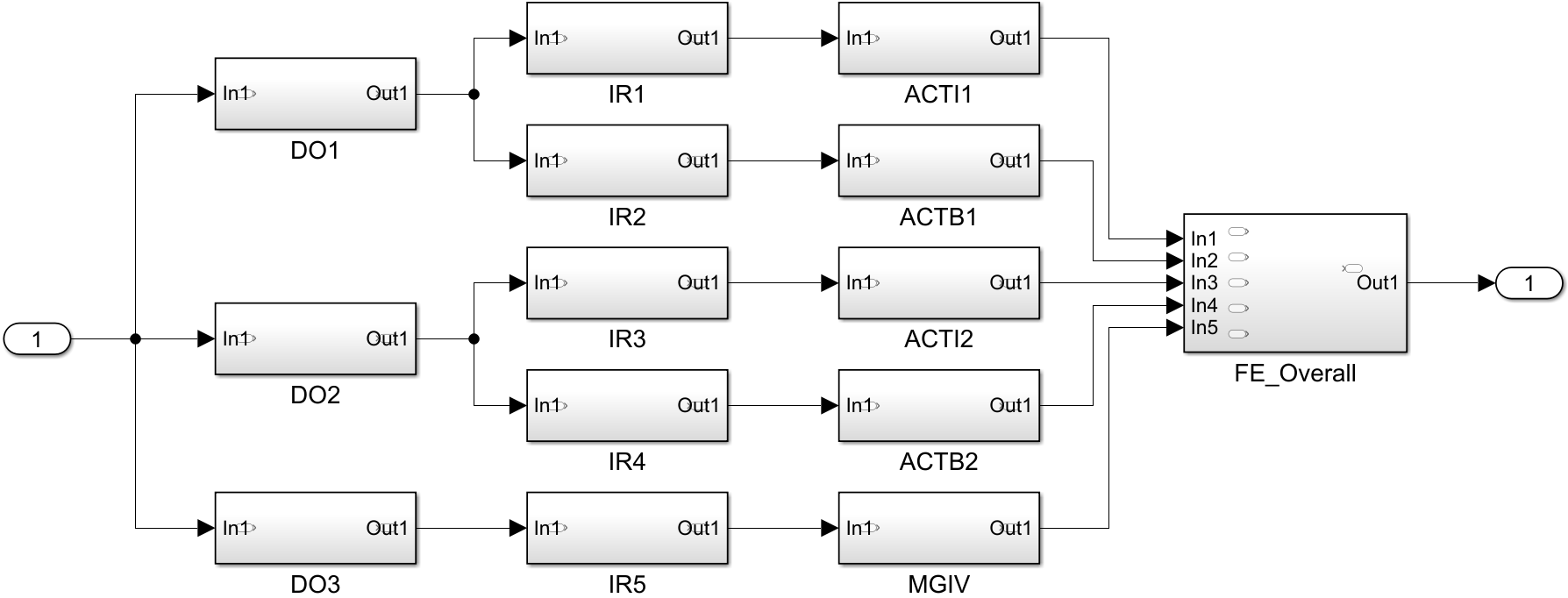}}
\caption{HiP-HOPS models in MATLAB}
\label{Fig_MATLAB2}
\end{figure}

We created a HiP-HOPS model in its user interface in the MATLAB environment. The interfacing between FMR and this model was done through an XML file exchange. The model was structured in two levels of hierarchy: system level (Figure \ref{Fig_MATLAB}), and component level (Figure \ref{Fig_MATLAB1} for the final elements). The DU and ST failures of SIS are the result of failures in SIS\_Inputs, SIS\_CPU or SIS\_FinalElements. The component failure modes of the latter two blocks are manually implemented in MATLAB whereas the failure modes of the SIS\_Inputs block are generated in FMR and automatically exported to a suitable data format in HiP-HOPS.

In Figure \ref{Fig_MATLAB}, the SIS\_CPU block consists of two failure components: the CPU module, and the communication link between CPU and input/output modules. A DU (or ST) failure of either of these two components can result in the failure of SIS\_CPU block and thus the overall failure of SIS. The SIS\_FinalElement block models the failure of DO modules, interposing relays and the valves. As shown in detail in Figure \ref{Fig_MATLAB1}, DO1 is shared between IBV1 and MBV1, and DO2 between IBV2 and MBV2. The main gas isolation valve (MGIV) is separately connected to DO3. The failure combinations for final elements are defined as follows: \CompCode{Out1.DU=(In1.DU AND In3.DU) OR (In2.DU AND In4.DU)} and \CompCode{Out1.ST=(In1.ST OR In.2-ST OR In3.ST OR In4.ST OR In5.ST)}. The analysis in HiP-HOPS produced the MCSs for all SIS subsystems. The CPU and final elements (FE) parts are shown in Tables \ref{Table_HIPHOPS_ST} and \ref{Table_HIPHOPS_DU}. The MCSs of inputs were the same as Tables \ref{Table_FMR_DU} and \ref{Table_FMR_ST}. 

\begin{minipage}{1.5in}
\begin{longtable}[!h]{|M{0.5cm}|M{2cm}|M{1.5cm}|N|}
\hline \multicolumn{1}{|c|}{\TE{No.}}&\multicolumn{1}{c|}{\TE{Min Cut Set}}&\multicolumn{1}{c|}{\TE{Frequency}}\\ 
\specialrule{0.2em}{0.0em}{0.0em} 
\TE{1} & \TE{CPU.CPUST} & \TE{5.00E-09}\\ 
\TE{2} & \TE{Comm.CommST} & \TE{1.00E-09}\\ 
\TE{3} & \TE{ACTB1.ACTBST} & \TE{8.00E-07}\\ 
\TE{4} & \TE{ACTB2.ACTBST} & \TE{8.00E-07}\\ 
\TE{5} & \TE{ACTI1.ACTIST} & \TE{8.00E-07}\\ 
\TE{6} & \TE{ACTI2.ACTIST} & \TE{8.00E-07}\\ 
\TE{7} & \TE{DO1.DOST} & \TE{1.00E-09}\\ 
\TE{8} & \TE{DO2.DOST} & \TE{1.00E-09}\\ 
\TE{9} & \TE{DO3.DOST} & \TE{1.00E-09}\\ 
\TE{10} & \TE{IR1.IRST} & \TE{4.00E-08}\\ 
\TE{11} & \TE{IR2.IRST} & \TE{4.00E-08}\\ 
\TE{12} & \TE{IR3.IRST} & \TE{4.00E-08}\\ 
\TE{13} & \TE{IR4.IRST} & \TE{4.00E-08}\\ 
\TE{14} & \TE{IR5.IRST} & \TE{4.00E-08}\\ 
\TE{15} & \TE{MGIV.ACTMST} & \TE{1.20E-06}\\ 
\hline 
\caption{CPU and FE MCSs for SIF ST}\label{Table_HIPHOPS_ST}
\end{longtable}
\end{minipage}\qquad\quad
\begin{minipage}{2.5in}
\begin{longtable}[!h]{|M{0.5cm}|M{2cm}|M{2cm}|M{1.5cm}|N|}
\hline \multicolumn{1}{|c|}{\TE{No.}}&\multicolumn{2}{c|}{\TE{Min Cut Set}}&\multicolumn{1}{c|}{\TE{Unavailability}}\\ 
\specialrule{0.2em}{0.0em}{0.0em} 
\TE{1} & \TE{CPU.CPUDU} & \TE{} & \TE{1.90E-04}\\[-0.2em] 
\TE{2} & \TE{Comm.CommDU} & \TE{} & \TE{1.00E-05}\\[-0.2em] 
\TE{3} & \TE{ACTB1.ACTBDU} & \TE{ACTB2.ACTBDU} & \TE{1.70E-04}\\[-0.2em] 
\TE{4} & \TE{ACTB1.ACTBDU} & \TE{DO2.DODU} & \TE{1.14E-07}\\[-0.2em] 
\TE{5} & \TE{ACTB1.ACTBDU} & \TE{IR4.IRDU} & \TE{6.86E-06}\\[-0.2em] 
\TE{6} & \TE{ACTB2.ACTBDU} & \TE{DO1.DODU} & \TE{1.14E-07}\\[-0.2em] 
\TE{7} & \TE{ACTB2.ACTBDU} & \TE{IR2.IRDU} & \TE{6.86E-06}\\[-0.2em] 
\TE{8} & \TE{ACTI1.ACTIDU} & \TE{ACTI2.ACTIDU} & \TE{1.70E-04}\\[-0.2em] 
\TE{9} & \TE{ACTI1.ACTIDU} & \TE{DO2.DODU} & \TE{1.14E-07}\\[-0.2em] 
\TE{10} & \TE{ACTI1.ACTIDU} & \TE{IR3.IRDU} & \TE{6.86E-06}\\[-0.2em] 
\TE{11} & \TE{ACTI2.ACTIDU} & \TE{DO1.DODU} & \TE{1.14E-07}\\[-0.2em] 
\TE{12} & \TE{ACTI2.ACTIDU} & \TE{IR1.IRDU} & \TE{6.86E-06}\\[-0.2em] 
\TE{13} & \TE{DO1.DODU} & \TE{DO2.DODU} & \TE{7.69E-11}\\[-0.2em] 
\TE{14} & \TE{DO1.DODU} & \TE{IR3.IRDU} & \TE{4.61E-09}\\[-0.2em] 
\TE{15} & \TE{DO1.DODU} & \TE{IR4.IRDU} & \TE{4.61E-09}\\[-0.2em] 
\TE{16} & \TE{DO2.DODU} & \TE{IR1.IRDU} & \TE{4.61E-09}\\[-0.2em] 
\TE{17} & \TE{DO2.DODU} & \TE{IR2.IRDU} & \TE{4.61E-09}\\[-0.2em] 
\TE{18} & \TE{IR1.IRDU} & \TE{IR3.IRDU} & \TE{2.77E-07}\\[-0.2em] 
\TE{19} & \TE{IR2.IRDU} & \TE{IR4.IRDU} & \TE{2.77E-07}\\ 
\hline 
\caption{CPU and FE MCSs for SIF DU}\label{Table_HIPHOPS_DU}
\end{longtable}
\end{minipage}

\medskip

The SIS inputs failure data were transferred automatically from FMR whereas the failure data for CPU and final elements were manually annotated in HiP-HOPS. We used the manufacturer's data as shown in Table \ref{Table_FailureData}. 

\begin{longtable}[!h]{|M{3cm}|M{2.7cm}|M{2.7cm}|M{2.7cm}|N|}
\hline \multicolumn{1}{|c|}{\TE{Component}}&\multicolumn{1}{c|}{\TE{Dormant (DU, SU), p.h.}}&\multicolumn{1}{c|}{\TE{Rate (DD, SD), p.h.}}&\multicolumn{1}{c|}{\TE{Fixed (PFD\textsubscript{avg})}}\\ 
\specialrule{0.2em}{0.0em}{0.0em} 
\TE{SIS CPU} & \TE{} & \TE{5.00E-9} & \TE{1.90E-4}\\& & & \\[-1.2em] \hline & & & \\[-1.2em] 
\TE{SIS Comm} & \TE{} & \TE{1.00E-9} & \TE{1.00E-5}\\& & & \\[-1.2em] \hline & & & \\[-1.2em] 
\TE{Digital Output Module} & \TE{1.00E-9} & \TE{1.00E-9} & \TE{}\\& & & \\[-1.2em] \hline & & & \\[-1.2em] 
\TE{Interposing Relay} & \TE{6.00E-8} & \TE{4.00E-8} & \TE{}\\& & & \\[-1.2em] \hline & & & \\[-1.2em] 
\TE{Igniter/Burner Block Valve} & \TE{1.50E-6} & \TE{8.00E-7} & \TE{}\\& & & \\[-1.2em] \hline & & & \\[-1.2em] 
\TE{Main Gas Valve} & \TE{} & \TE{1.20E-6} & \TE{}\\ 
\hline 
\caption{SIS component failure data}\label{Table_FailureData}
\end{longtable}

With the same MTTR=8 hours and Risk Assessment Time and Proof Test Interval of 2 years, the overall model, including the imported FMR part, was analyzed in HiP-HOPS and the following results were obtained for the overall SIF: $Q_{DU}=1.88E-03$ and $W_{ST}= 4.75E-06$. The results generated by HiP-HOPS matched up the ones of our reference model in FaultTree+. 
 
\section{Integration with CFT}\label{Sec_model3}

A CFT is a Boolean model associated to system development elements such as components \cite{Ref_143}. It has the same expressive power as classic fault trees and, likewise, it is used to model failure behavior of safety-critical systems.

In CFTs, every component is represented by a CFT element. Each element has its own in-ports and out-ports that are used to express propagation of failure modes through the tree. Similar to classic fault trees, the internal failure behavior that influences the output failure modes is modeled by Boolean gates. 

The main difference between the two methods is that unlike classic fault trees, CFTs can have multiple top events (e.g. both the DU and ST modes) within the same model. Thus, the tree structure in CFT is extended towards a Directed Acyclic Graph. This eliminates the need for artificial splitting of common cause failure into multiple repeated events, and makes it possible to have more than one path to start from the same basic event or sub-tree. 

\begin{figure}[!ht]
\centering
\includegraphics[width=6cm]{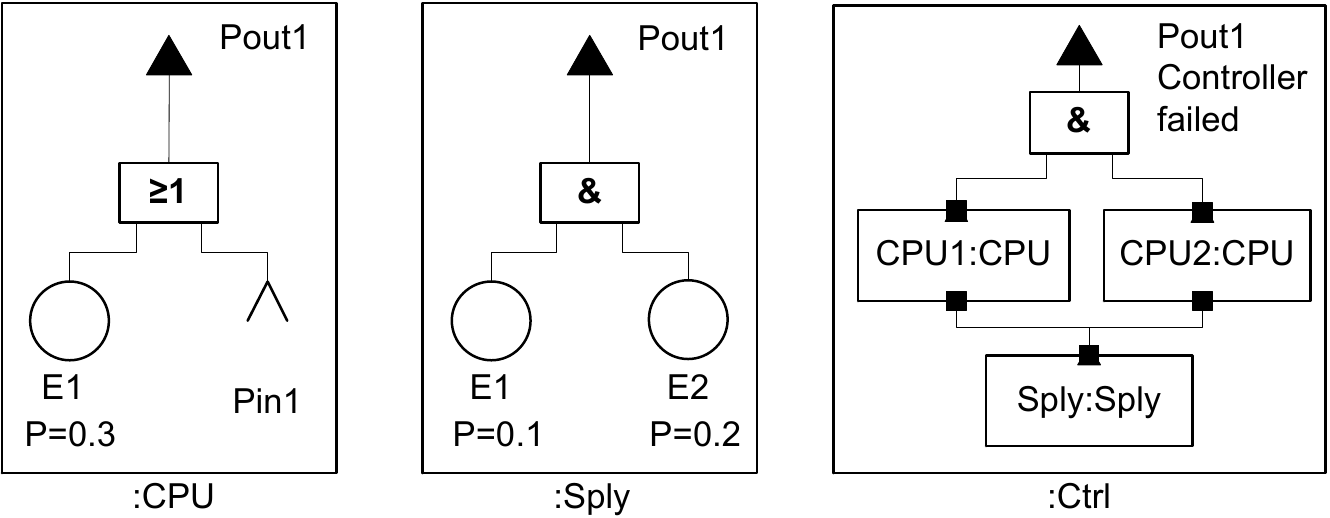}
\caption{Example of a simple CFT}
\label{fig:CFT_Introduction_Example}
\end{figure}

A small example of a CFT was presented in \cite{Ref_205} (see Figure~\ref{fig:CFT_Introduction_Example}). 
The example shows an exemplary controller system \emph{Ctrl}, including two redundant \emph{CPU}s (i.e.~two instances of the same component type) and one common power supply \emph{Sply}, which would be a repeated event in traditional fault tree. The controller is unavailable if both CPUs are in the ``failed" state. The inner fault tree of the CPU is modeled as a type. Since the CPUs are identical, they only have to be modeled once and then instantiated twice in the main model. The failure of a CPU can be caused by some inner basic event E1, or by an external failure which is connected via the in-port. As both causes result in a CPU failure, they are joined via an OR gate. The power supply module is modeled as another type. In this example the power supply is in its ``failed" state if both basic failures E1 and E2 occur. Hence, instead of a single large fault tree, the CFT model consists of small, reusable and easy-to-review components.

\begin{figure}[!ht]
\includegraphics[scale=0.19]{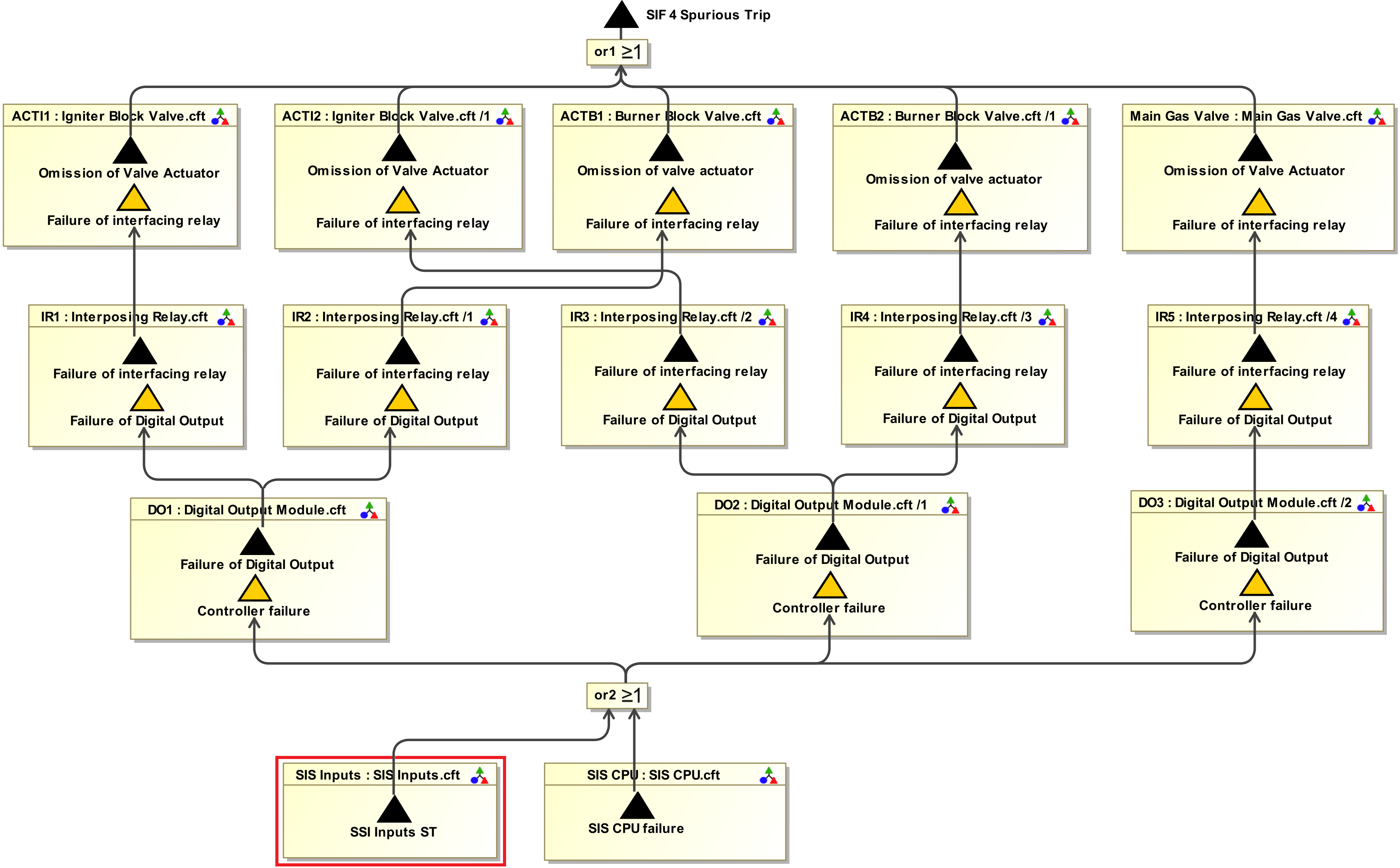}
\centering
\caption{CFT model for ST failure}
\label{Fig_CFT_ST}
\end{figure}

Similar to the HiP-HOPS experiment, we implemented an automatic data link  between FMR and CFT. The list of MCSs, including the model types and failure rates of basic events were exported in CSV format to the CFT tool, where a new add-on script would read the data and compose a CFT element for SIS Inputs. The rest of the modeling, i.e. for CPU and final elements, was implemented manually in the CFT tool. Figure \ref{Fig_CFT_ST} shows the CFT model for ST failure. The highlighted box represents the SIS Inputs, to which the FMs are imported from FMR. A similar model was developed for analyzing DU failure.

CFT analysis produced the same list of MCSs as in HiP-HOPS and FaultTree+. Using the model types and failure rates of basic events shown in Table \ref{Table_CFT}, the tool generated the following quantitative results:
\begin{itemize}
\item Average failure probability in DU mode $Q_{DU} = 1.0E-3$ 
\item Mean failure rate in ST mode $W_{ST}$ = 4.61E-6 p.h. 
\end{itemize}
It is apparent that the CFT results differ from what we saw in the previous section. The main reason is that the approximation methods used for calculating the impact of common basic events are different in different tools. The quantitative results presented in the previous two sections used Eqs. \ref{Eq_MCSQW} and \ref{Eq_TEQW}, whereas CFT is based on the Siemens' internal tool ZUSIM, which uses the approach described in \cite{Ref_210,Ref_211}. By changing the settings of approximation method, in FaultTree+ for instance, we could observe narrower gaps between the results.  

\begin{longtable}[!h]{|M{3.5cm}|M{3.5cm}|M{3.5cm}|}
\hline \multicolumn{1}{|c|}{\TE{Basic Event}}&\multicolumn{1}{c|}{\TE{DU}}&\multicolumn{1}{c|}{\TE{ST}}\\ 
\specialrule{0.2em}{0.0em}{0.0em} 
\TE{SIS CPU} & \TE{Probability = 1.9E-4} & \TE{$\lambda$ = 5.0E-9}\\ & & \\[-1.2em] \hline & & \\[-1.2em] 
\TE{SIS Comm} & \TE{Probability = 1.0E-5} & \TE{$\lambda$ = 1.0E-9}\\ & & \\[-1.2em] \hline & & \\[-1.2em] 
\TE{Digital Output Module} & \TE{$\lambda$ = 1.0E-9} & \TE{$\lambda$ = 1.0E-9}\\ & & \\[-1.2em] \hline & & \\[-1.2em] 
\TE{Interposing Relay} & \TE{$\lambda$ = 6.0E-8} & \TE{$\lambda$ = 4.0E-8}\\ & & \\[-1.2em] \hline & & \\[-1.2em] 
\TE{Igniter/Burner Block Valve} & \TE{$\lambda$ = 1.5E-6} & \TE{$\lambda$ = 8.0E-7}\\ & & \\[-1.2em] \hline & & \\[-1.2em] 
\TE{Main Gas Valve} & \TE{} & \TE{$\lambda$ = 1.2E-6}\\ & & \\[-1.2em] \hline & & \\[-1.2em] 
\TE{Input "healthy"} & \TE{Probability = 0.999} & \TE{Probability = 0.999}\\ & & \\[-1.2em] \hline & & \\[-1.2em] 
\TE{Input "faulty"} & \TE{Probability = 2.0E-6} & \TE{Probability = 2.0E-6}\\ & & \\[-1.2em] \hline & & \\[-1.2em] 
\TE{Input "healthy \& higher/lower"} & \TE{Probability = 4.379E-4} & \TE{Probability = 4.379E-4}\\ 
\hline 
\caption{Types and rates used for CFT modeling}\label{Table_CFT}
\end{longtable}

\section{Discussion}\label{Sec_discussion}

All SIS components are important, but no SIS analysis can be complete without including the behavior of its data-processing, decision-making program. The problem is that such analyses can be painstaking and time-consuming in complex systems, and when done by hand the results will still be susceptible to human error. Consequently, current SIS analyses often lack this critical part, and use simplifications and assumptions instead, which can lead to unreliable results. FMR solves this problem by automating the process and by studying the exact program that the SIS would execute. However, FMR's visibility is understandably limited to what influences the program. Hence, an integration with other generic FTA tools can provide a complete coverage. As such, each tool would still do what they are good at while the integration achieves an inclusive outcome. 

We demonstrated through a case study how such integration can be implemented in practice. We chose HiP-HOPS and CFT as two different examples, both with proven records in other applications, and both from industries other than process. The fact is that FMR can integrate with any FTA-based method that allows standard file formats, e.g. XML and CSV, for data exchange. 

The underlying question in FMR is: given a resultant deviation at the output and given the actual system program, what are the possible causing deviations at the input. This is obviously different to FTA, where we ``know'' the failure behavior of a system and we build a model (fault tree) to summarize our understandings. FMR is rather a failure identification method, one that can be used in failure modeling applications. Nonetheless, FMR shares a key aspect with FTA-based modeling methods such as HiP-HOPS and CFT: a component-based approach in failure analysis. Compared to conventional FTA, component-based methods provide better visibility to failure behavior of systems. Traditional fault trees become visually hard to navigate as the model size grows. Hierarchical, topographic models, such as the one in Figure \ref{Fig_MATLAB2}, offer an easier and more transparent understanding of the relationship between subsystems and components at various levels, which enhances the qualitative analysis of safety systems.

Furthermore, a safety analysis can be improved by selecting the ``right" method of calculation. There are different approximation methods, referred to by different names, including Rare Event (RE), Inclusion-Exclusion (IE), Esary-Proschan (EP) and Cross-Product, depending on which the results may vary. This may in turn lead to requiring structural changes in the SIS design, if the reliability targets are not met \cite{Ref_190}. See Table \ref{Table_FTResults} as an example from FaultTree+; the results would change if we chose a different method in our case study. 

\begin{longtable}[!h]{|M{3cm}|M{2.7cm}|M{2.7cm}|M{2.7cm}|N|}
\hline \multicolumn{1}{|c|}{\TE{Calculation}}&\multicolumn{1}{c|}{\TE{Default}}&\multicolumn{1}{c|}{\TE{Esary-Proschan}}&\multicolumn{1}{c|}{\TE{Rare Event}}\\ 
\specialrule{0.2em}{0.0em}{0.0em} 
\TE{DU  unavailability} & \TE{1.00E-3} & \TE{1.88E-3} & \TE{1.88E-3}\\& & & \\[-1.2em] \hline & & & \\[-1.2em] 
\TE{ST  frequency} & \TE{4.75E-6} & \TE{4.75E-6} & \TE{4.76E-6}\\ 
\hline 
\caption{FaultTree+ calculations for different approximation methods}\label{Table_FTResults}
\end{longtable}

Among various approximation methods, we use the EP \cite{Ref_207} method for FMR, as it is more conservative than the IE formula itself but less of the one of RE \cite{Ref_176}. The same selection was set in HiP-HOPS and FaultTree+ so that we could compare the results. A different calculation method as described in \cite{Ref_210} is used to analyze CFTs. Here, we set the selection in FaultTree+ to its default upper bound approximation so we could verify the CFT results. 

Modeling of CPU and final elements (FE) in HiP-HOPS and CFT was done manually. However, the effort required for modeling these parts is not comparable to analyzing the program, which was done automatically. Our case study SIS implemented 34 SIFs. Considering an average of 30 MCSs for each SIF (our case study SIF had 45), the analyst would need to identify 1020 MCSs for SIS inputs. The number of MCSs in CPU and FE parts combined was only 34, which is almost 3\% of the overall. This is because the CPU and FE parts are common between all those 34 SIFs, and thus they are modeled once; but the inputs to each SIF need a separate model on its own. Besides, the level of complexity in CPU and FE failures is considerably lower than those in a program.  

\section{Conclusion}\label{Sec_conclusion}
 
We demonstrated two practical examples of integrating FMR with model-based methods HiP-HOPS and CFT. The purpose of this study was to experience comprehensive safety analyses, that included the impact of an SIS program in precise detail. In this project, FMR was used to analyze the SIS input subsystem while the random failure of logic solver and final elements were modeled in the other tools. Add-on codes were developed in each individual tool to enable automated data interfacing while the analysis methods in each tool remained unchanged. In parallel, we created a separate model in FaultTree+, to compare and verify the results of our own models with one same reference. 

The main achievement of this study was showing how SIS programs can be included in safety analyses and how integrating between FMR and other FTA-based tools can help overcome modeling challenges associated with programs. Benefits of the integration include enhanced model accuracy, expanded modeling coverage, reduced modeling effort and improved analysis performance. The success of this project provided a platform for improved safety analyses in the process industry. Future research work will include expanding the interfacing features of the FMR tool, extending FMR to analyzing failure modes of system parameters, and adapting the method for modeling generic systems. In the meantime, we are in the process of publishing a formal proof for the theoretical foundations of FMR to better support its use in safety-related analyses.  

\bibliographystyle{splncs04}
\bibliography{References}

\begin{thebibliography}{10}
\providecommand{\url}[1]{\texttt{#1}}
\providecommand{\urlprefix}{URL }
\providecommand{\doi}[1]{https://doi.org/#1}

\bibitem{Ref_207}
Henley, E.J., Kumamoto, H.: Probabilistic risk assessment and management for
  engineers and scientists. IEEE Press (2nd Edition)  (1996)

\bibitem{Ref_182}
IEC: {IEC 61025: Fault tree analysis (FTA)}  (2006)

\bibitem{Ref_188}
IEC: {IEC 61508: Functional safety of electrical/electronic/programmable
  electronic safety related systems - Part 6: Guidelines on the application of
  IEC 61508-2 and IEC 61508-3} (2010)

\bibitem{Ref_203}
IEC: {Programmable controllers - Part 3: Programming languages}  (2013)

\bibitem{Ref_190}
IEC: {Functional safety-Safety instrumented systems for the process industry
  sector - Part 1: Framework, definitions, system, hardware and application
  programming requirements}  (2016)

\bibitem{Ref_196}
ISA: {ISA-TR84.00.02-2015, Safety Integrity Level (SIL) Verification of Safety
  Instrumented Functions}  (2015)

\bibitem{Ref_197}
Jahanian, H.: {Generalizing PFD formulas of IEC 61508 for KooN configurations}.
  ISA transactions  \textbf{55},  168--174 (2015)

\bibitem{Ref_200}
Jahanian, H.: Failure mode reasoning. In: 2019 4th International Conference on
  System Reliability and Safety (ICSRS). pp. 295--303. IEEE (2019)

\bibitem{Ref_143}
Kaiser, B., Liggesmeyer, P., M{\"a}ckel, O.: A new component concept for fault
  trees. In: Proceedings of the 8th Australian workshop on Safety critical
  systems and software-Volume 33. pp. 37--46. Australian Computer Society, Inc.
  (2003)

\bibitem{Ref_205}
Kaiser, B., Schneider, D., Adler, R., Domis, D., M{\"o}hrle, F., Berres, A.,
  Zeller, M., H{\"o}fig, K., Rothfelder, M.: Advances in component fault trees.
  In: Proc. of ESREL (2018)

\bibitem{Ref_71}
Papadopoulos, Y., McDermid, J., Sasse, R., Heiner, G.: Analysis and synthesis
  of the behaviour of complex programmable electronic systems in conditions of
  failure. Reliability Engineering \& System Safety  \textbf{71}(3),  229--247
  (2001)

\bibitem{Ref_222}
Papadopoulos, Y., Walker, M., Parker, D., Sharvia, S., Bottaci, L., Kabir, S.,
  Azevedo, L., Sorokos, I.: A synthesis of logic and bio-inspired techniques in
  the design of dependable systems. Annual Reviews in Control  \textbf{41},
  170--182 (2016)

\bibitem{Ref_209}
Parker, D., Walker, M., Papadopoulos, Y.: Model-Based Functional Safety
  Analysis and Architecture Optimisation, pp. 79--92. IGI Global (2013)

\bibitem{Ref_194}
Rausand, M.: Reliability of safety-critical systems. John Wiley\&Sons  (2014)

\bibitem{Ref_211}
Stecher, K.: Fault tree analysis, taking into account causes of common mode
  failures. Siemens Forschungs- und Entwicklungsberichte  (1984)

\bibitem{Ref_210}
Stecher, K.: Evaluation of large fault-trees with repeated events using an
  efficient bottom-up algorithm. IEEE transactions on reliability
  \textbf{35}(1),  51--58 (1986)

\bibitem{Ref_176}
Vesely, W.E., Goldberg, F.F., Roberts, N.H., Haasl, D.F.: Fault Tree Handbook
  (NUREG-0492). {US Nuclear Regulatory Commission} (1981)

\end{thebibliography}

\end{document}